\documentclass[useAMS,usenatbib,usegraphicx]{mn2e}
\usepackage{psfrag}
\newcommand{\jhk}{\ensuremath{JHK\!s}}
\newcommand{\lm}{\ensuremath{L'M'}}
\newcommand{\jhklm}{\ensuremath{JHK\!sL'M'}}
\title[The centre of the large circumstellar disc in M\,17]{Probing the
centre of the large circumstellar disc in M\,17\thanks{Based on observations collected at the European Southern Observatory, Chile., project nos. 73.C-0170, 73.C-0183 and 75.C-0852.}}

\author[M. Nielbock et al.]{M.~Nielbock,$^1$\thanks{E-mail: nielbock@mpia.de}
        R.~Chini,$^2$
        V.~H.~Hoffmeister,$^2$
        D.~E.~A.~N\"urnberger,$^3$
        \newauthor
        C.~M.~Scheyda$^2$
        and J.~Steinacker$^{1,4}$\\
$^1$Max-Planck-Institut f\"ur Astronomie,
    K\"onigstuhl 17, D-69117 Heidelberg, Germany\\
$^2$Astronomisches Institut der Ruhr-Universit\"at Bochum,
Universit\"atsstra{\ss}e 150/NA 7, D-44780 Bochum, Germany\\
$^3$European Southern Observatory, Alonso de Cordova 3107, Casilla
19001, Santiago 19, Chile\\
$^4$Astronomisches Rechen-Institut, Zentrum f\"ur Astronomie
Heidelberg, M\"onchhofstr. 12-14, D-69120 Heidelberg, Germany}
\begin{document}

\date{Received / Accepted}
\pagerange{\pageref{firstpage}--\pageref{lastpage}} \pubyear{2008}

\maketitle

\label{firstpage}

\begin{abstract}
We investigated the nature of the hitherto unresolved elliptical infrared
emission in the centre of the $\sim$20\,000\,AU disc silhouette in M\,17. We
combined high-resolution {\jhklm}~band imaging carried out with NAOS/CONICA at
the VLT with [Fe\,{\sc ii}] narrow~band imaging using SOFI at the NTT. The
analysis is supported by Spitzer/GLIMPSE archival data and by already published
SINFONI/VLT Integral Field Spectroscopy data. For the first time, we resolve the
elongated central infrared emission into a point-source and a jet-like feature
that extends to the northeast in the opposite direction of the recently
discovered collimated H$_2$ jet. They are both orientated almost perpendicular
to the disc plane. In addition, our images reveal a curved southwestern emission
nebula whose morphology resembles that of the previously detected northeastern
one. Both nebulae are located at a distance of 1500\,AU from the disc centre. We
describe the infrared point-source in terms of a protostar that is embedded in
circumstellar material producing a visual extinction of $60 \leq A_V \leq 82$.
The observed $K\!s$~band magnitude is equivalent to a stellar mass range of
$2.8\,M_{\sun} \leq M_\star \leq 8\,M_{\sun}$ adopting conversions for a
main-sequence star. Altogether, we suggest that the large M\,17 accretion disc
is forming an intermediate to high-mass protostar. Part of the accreted material
is expelled through a symmetric bipolar jet/outflow.
\end{abstract}

\section{Introduction}
The processes forming high-mass stars are still not well understood. Although
recent theoretical models require the presence of and the accretion from a
circumstellar disc (\citealt{bonnell01}; \citealt*{yorke02};
\citealt*{krumholz07disks}), the direct observation of such a configuration is
still missing.

During a systematic infrared study of the young star cluster NGC~6618 inside the
H\,{\sc ii} region M\,17 (Omega Nebula), \citet{chini04nat} discovered an opaque
silhouette, shaped like a flared disc at $JHK$ with a diameter of
$\sim$20\,000\,AU against the bright background of the H\,{\sc ii} region. It is
associated with an optically visible hourglass-shaped nebula perpendicular to
the silhouette plane. The optical spectrum of the nebula exhibits emission lines
with blue-shifted absorption features indicating disc accretion.

\citet{steinacker06} successfully modelled the optical depth and the photon
scattering of the silhouette at $2.2\,\umu$m and obtained a large circumstellar
disc seen at an inclination angle of $12^\circ$ (almost edge-on). They found
disc masses between 0.02 and $5\,M_{\sun}$ depending on the assumed dust model
and the distance. The mass range is in good agreement with the distance-scaled
values by \citet{sako05} derived from their NIR (near infrared) extinction map.
The distance has recently been determined to $2.1 \pm 0.2$\,kpc on the basis of
about 50 spectroscopically classified early-type stars \citep{hoffmeister08}
ruling out earlier estimates below 2\,kpc that treated several unresolved
high-mass binaries as single stars. The new spectro-photometric distance agrees
well with previous studies based on multi-colour photometry
\citep*[$2.2\pm0.2$\,kpc;][]{chini80} and radio data
\citep[$2.4\pm^{0.4}_{0.5}$\,kpc;][]{russeil03}.

Independent of its mass, the M\,17 silhouette is the largest disc found to date that is exhibiting accretion activity. In fact, \citet{nuernberger07} discovered a H$_2$ jet emerging from the disc centre, also suggesting ongoing accretion.

At the disc centre, \citet{chini04nat} found an elliptical $K$~band emission of $240 \times 450$\,AU$^2$. Likewise, \citet{sako05} describe the central object as an elongated compact infrared source. A faint emission of unknown nature is also visible in their [Ne\,{\sc ii}] narrow~band MIR (mid infrared) image at $12.8\,\umu$m.

Assuming that the central $K$~band flux found by \citet{chini04nat} originates
from a single star at a distance of 2.1\,kpc, the stellar mass may range between
a few and $45\,M_{\sun}$, depending on the extinction model
\citep{steinacker06}. \citet{sako05} explained their results in terms of an
intermediate-mass object with a stellar mass of $2.5$ to $8\,M_{\sun}$. Their
analysis was part of a larger survey of candidate silhouette objects in M\,17
\citep{ito08}.

\begin{figure*}
\centering
\resizebox{\hsize}{!}{\includegraphics[angle=270]{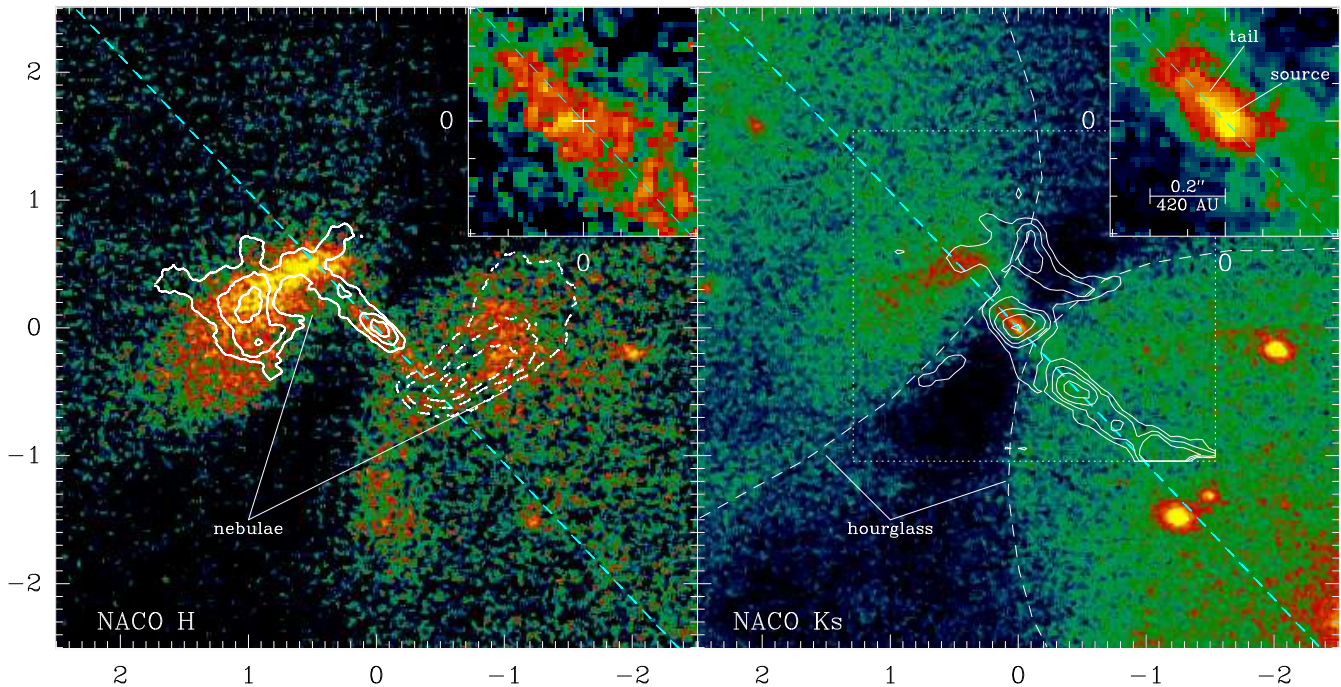}}
\caption{\label{f:hk_image} Infrared images of the inner region
($5\arcsec \times 5\arcsec$) of the M\,17 disc silhouette centred on
the newly detected point-source at R.A. = $18^{\rm h}20^{\rm m}26\fs191$
and Dec. = $-16^\circ12'10\farcs46$ (J2000). The inserts enlarge the central $0\farcs6\times0\farcs6$.
Left: $H$ image showing the symmetric bipolar nebula. The central point-source
is not detected. To visualise the symmetry of the two nebulae, we have
superimposed their contours after rotating them by $180^\circ$ around the
central point-source which we define as the symmetry centre; solid contours
refer to the southwestern lobe, dashed contours represent the northeastern
one.
Right: $K\!s$ image resolving the central NIR emission into a point-source and a
fainter tail extending to the northeast. The hourglass structure above and below
the disc hosts the bipolar nebula, of which the southwestern lobe is not
detected. The contours denote the continuum-free H$_2$ $1-0$\,S(1) line emission
at $\lambda = 2.12\,\umu$m as published by \citet{nuernberger07}; the dotted box
delineates the extent of the H$_2$ image. The dashed blue lines mark the
suggested jet axis as determined by the orientation of the detected tail with
respect to the point-source.}
\end{figure*}

So far, both the mass and the temperature of the forming star within this
spectacular disc remain under discussion. In this work, we present new data that
provide further insight into the nature of the central protostellar object and
constrain the possible mass range. We also discuss the detection of a possible
counter-jet headed in the opposite direction of the H$_2$ jet and a curved
southwestern emission nebula that is geometrically symmetric to the northeastern
one.

\section{Observations and reduction}
The narrow~band imaging was done with SOFI \citep{sofi} at the ESO NTT on La
Silla, Chile, in April 2004. The [Fe\,{\sc ii}] line at $\lambda=1.64\,\umu$m
was probed with a narrow band filter centred on $\lambda=1.644\pm0.025\,\umu$m.
In order to subtract the continuum from the spectral line data, we obtained
images in an adjacent narrow band filter centred on $\lambda=1.71\,\umu$m. The
seeing limited spatial resolution is 0\farcs9 (FWHM), the pixel size is
0\farcs288.

The adaptive optics {\jhklm} imaging was carried out in July and
August 2005 with NAOS/CONICA \citep{naco1,naco2} at the ESO VLT. We
used the CONICA S13 lens for the {\jhk} measurements providing a
pixel resolution of $0\farcs013$ and a field size of $14\arcsec
\times 14\arcsec$. The {\lm} imaging was done with the L27 lens
resulting in a pixel size of $0\farcs027$. While for the $L'$~band
the corresponding FOV was $28\arcsec \times 28\arcsec$, $M'$~band
measurements only use the central $512\times 512$ pixels equivalent
to a field size of $14\arcsec \times 14\arcsec$. The effective
spatial resolution is $0\farcs1$ (FWHM). The photometric errors vary
between $0.1$\,mag and $0.4$\,mag.

The Spitzer post-BCD data were taken from the GLIMPSE survey archive
\citep{spitzer-glimpse} based on an imaging campaign of the galactic
plane using the IRAC camera \citep{spitzer-irac} on board the
Spitzer infrared satellite \citep{spitzer}.

All data were reduced and analysed with IRAF and MOPSI; the NACO $H$ and
$K\!s$~band images are shown in Fig.~\ref{f:hk_image}. Among the NIR AO data,
only the $K\!s$~band image shows a point-source; at $J$ and $H$ the emission is
diffuse and extended. It was not detected at all in the $L'M'$ bands. Therefore,
establishing a meaningful SED (spectral energy distribution) for the embedded
source is difficult. We were able to produce adequate PSF models in the NACO
$HK\!sL'M'$ bands by selecting 6 to 10 sufficiently bright stars also detected
on the same frame for the subsequent PSF photometry. Within the errors, the
$K\!s$ PSF fit to the central point-source appeared circular. The surrounding
fuzzy and somewhat scattered emission can be entirely attributed to the central
ridge introducing a background pattern similar to what is visible in the $H$
band. Due to the absence of sufficiently bright PSF model stars in the NACO
$J$~band, we had to perform aperture photometry. An aperture correction was
applied to all the measurements. The results of the photometry are listed in
Table~\ref{t:phot}.

Astrometric calibration of all but the NACO data is based on the
2MASS database. Since there are no 2MASS sources on the NACO frames,
we used stars detected with ISAAC \citep{chini04} in the vicinity of
the disc for astrometry. We estimate the accuracy to be better than 0\farcs1.

\section{Results and discussion}
\subsection{The central point-source}
\label{s:protostar}

\begin{table}
\tabcolsep5pt
\caption{NACO multi-colour photometry of the central source.}
\label{t:phot}
\begin{tabular}{lcccc}
\hline
&\multicolumn{2}{c}{Point-source}
&\multicolumn{2}{c}{NE tail}\\
$\lambda_c$ [$\umu$m] & \multicolumn{1}{c}{$m$ [mag]} &
\multicolumn{1}{c}{$F$ [$\umu$Jy]}
& \multicolumn{1}{c}{$m$ [mag]} & \multicolumn{1}{c}{$F$ [$\umu$Jy]}\\
\hline
1.27 & $> 23.71$ & $< 0.5$ & $> 23.71$ & $< 0.5$  \\
1.66 & $> 20.74$ & $< 7.1$ & $> 20.74$ & $< 7.1$  \\
2.18 & $19.28\pm0.07$ & $12.9\pm0.9$ & $20.10\pm0.12$ & $6.1 \pm 0.7$ \\
3.80 & $> 13.67$ & $< 960$ & $> 13.67$ & $< 960$  \\
4.78 & $> 12.61$ & $< 1460$& $> 12.61$ & $< 1460$ \\
\hline
\end{tabular}
\end{table}

The $K\!s$~band image in Fig.~\ref{f:hk_image} for the first time resolves the
elliptical infrared emission at the disc centre into a point-source with a FWHM
of 0\farcs1 (210\,AU) and a tail that extends to the northeast. We regard the
point-source as a single object, although close high-mass binaries can be
separated by $\leq 0.25$\,AU \citep{krumholz07binaries}. The peak of the tail
emission is separated by 0\farcs08 (170\,AU) from the point-source, defining a
line that we interpret as the outflow axis (see Sect.~\ref{s:jet}). It appears
tilted by $\sim 15^\circ$ from the disc symmetry axis. In the $H$~band, we only
detect diffuse emission without any indication for a point-source.

The symmetry of the source-jet-disc configuration is striking. In particular the
link between the source and the jet seems obvious. The point-source bisects the
line between the peak emission of the northeastern nebula coinciding with a weak
H$_2$ feature and the first strong maximum of the jet to the southwest
(Fig.~\ref{f:hk_image}, right panel). If chosen as the symmetry centre of the
bipolar nebula, both wings match perfectly when rotated by $180^\circ$
(Fig.~\ref{f:hk_image}, left panel).

Assuming the point-source emission stems from a reddened stellar photosphere at
a distance of 2.1\,kpc, we can combine our photometry with the extinction
estimate of \citet{steinacker06}. They pursued the so far most sophisticated
approach to model the spatial distribution of the optical depth of the disc at
$2.2\,\umu$m up to an outer radius of 10\,000\,AU including photon scattering
effects. Their result of $\tau_{2.2\,\umu{\rm m}}=2.6$ ($A_V=27$) toward the
disc centre is a lower limit, because (a) the unresolved inner disc at radii of
$\leq 1000$\,AU were excluded from the model, and (b) the model accounts for the
optical depth of the disc alone without the influence of interstellar and
intra-cluster foreground extinction. The latter contributions are estimated by
analysing the NIR colours of stars in the vicinity of the disc derived from the
previously published ISAAC $JHK$~data \citep{chini04}. Typical extinction values
off the disc plane are about $A_V=35\pm5$. As a result, the sum of the
foreground and disc extinction toward the disc centre amounts to $A_V = 62 \pm
5$. \citet{ito08} find an absolute upper limit of $A_V=30$, but this is probably
biased by their detection limit. It contradicts previous estimates giving
extinction values of $A_V>40$ for individual stars in this area
\citep{chini98,jiang02,hoffmeister06}.

\begin{figure}
\centering
\resizebox{\hsize}{!}{\includegraphics[angle=270]{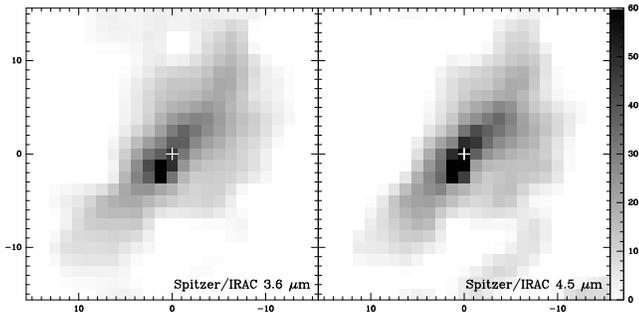}}
\caption{\label{f:extinction} Visual extinction maps of the central
$30''\times30''$ of the disc determined by applying the method of
\citet{siebenmorgen00} to Spitzer/IRAC data. The position of the
central point-source is marked with a white cross. The scale is
given in $A_V$. The extinction maximum is located $\sim 3''$
southeast of the disc centre.}
\end{figure}

We also estimated the extinction towards the disc centre independently by
creating an extinction map employing the method described by
\citet{siebenmorgen00}. It is based on a differential photometry of the
attenuating material (the disc) and a uniform background radiation (the H\,{\sc
ii} region). Considering a uniform foreground extinction across the extent of
the disc, the measured optical depth is purely due to the material inside the
disc, because other contributions along the line of sight like interstellar and
intra-cluster extinction or emission cancel out. Using Spitzer/GLIMPSE archival
data at $3.6$ and $4.5\,\umu$m, we determined the spatial distribution of the
optical depth by analysing the attenuation of the background emission
originating from the ionised gas. This approach is comparable to the NIR
extinction maps of \citet{sako05} and \citet{ito08}, but it has the advantage of
being less susceptible to scattering effects because of the larger wavelengths
covered by the Spitzer data. The optical depth estimates were converted to
visual extinction values (\citealt{flaherty07}; \citealt{roman07};
\citealt*{nielbock03}). In the resulting maps (Fig.~\ref{f:extinction}), the
extinction maximum is offset by $\sim 3''$ (6300\,AU) southeast from the central
point-source. Along the line of sight toward the disc centre, we obtain a total
extinction solely caused by the disc material of $A_V = 60 \pm 10$, half of
which is expected to be the disc extinction toward the centre. The true peak
extinction is probably even higher because of the spatial averaging resulting
from the limited resolution of the Spitzer data. This leads to a blending of
extinction and emission features resulting in a lower limit estimate of the
extinction. \citet{ito08} find similar displacements between the extinction
maximum of candidate silhouette objects and the suggested embedded YSO, which
they attribute either to a migration of the YSO or its off-centred formation.
However, it is still possible that those identified silhouette structures are
not necessarily related to the nearby YSOs. After adding the foreground
extinction, our results from the Spitzer extinction maps are consistent with the
estimate based on the NIR modelling.


Using the estimate of $A_V \sim 60\,$mag as a lower limit to the extinction, the
apparent $K\!s$ magnitude of the point-source is equivalent to an absolute
magnitude of a star having a spectral type of at least B8. In the case of a
main-sequence star, this would correspond to a stellar mass of $\sim
2.8\,M_{\sun}$ \citep*{blum00}. Although the reconstructed SINFONI H$_2$ line
image possesses a peak at the disc centre, the contribution of the H$_2$
emission to the continuum flux is negligible. The continuum flux measured in the
NACO $K\!s$ filter ($\Delta\lambda=0.35\,\umu$m) amounts to $2.9\times
10^{-18}$\,W\,m$^{-2}$; the H$_2$ $1-0$\,S(1) and H$_2$ $2-1$\,S(1) lines that
have been measured with SINFONI contain a continuum corrected line flux of
$2-8\times 10^{-20}$\,W\,m$^{-2}$. Thus, they contribute only $1-3$\% to the
continuum flux in the $K\!s$ filter. However, we cannot rule out any
additional contamination by excess emission of spatially unresolved
circumstellar material or accretion signatures close to the protostar.

\citet{sako05} present an image of this object obtained in the MIR using a
narrow-band filter centred on the [Ne\,{\sc ii}] line at 12.8\,$\umu$m. Although
a faint elongated emission is detected at the disc centre, it does not
contribute much to resolving the open question of the nature of the central
star. Since emission from this line is commonly observed for circumstellar discs
\citep[e.~g.][]{pascucci07}, it is not clear, how much of the detection is
caused by continuum emission.

A strict upper mass limit is difficult to determine from the available data.
\citet{sako05} argued that the missing radio continuum in the interferometric
radio map of \citet*{felli84} establishes an upper mass limit of $8\,M_{\sun}$.
However, this statement might not be valid, because the source is located in
projection onto an ionisation front exhibiting large scale free-free emission
that could cause confusion with any weak compact H\,{\sc ii} source like an
early-type star in the lower mass range.  Furthermore, a very young and just
developing ionised gas sphere (e.~g.~a hyper-compact H\,{\sc ii} region) would
be too small and too weak to be detected by its free-free emission
\citep{kurtz05}.

\begin{figure}
\centering
\psfrag{xaxis}{\LARGE$T_{\rm eff}$ [K]}
\psfrag{yaxis}{\LARGE$A_V$ [mag]}
\psfrag{b1}{\large B1V}
\psfrag{b2}{\large B2V}
\psfrag{b3}{\large B3V}
\psfrag{b5}{\large B5V}
\psfrag{b7}{\large B7V}
\psfrag{b9}{\large B9V}
\psfrag{ksm30}{\Large$K\!s-M=3.0$}
\psfrag{ksm40}{\Large$K\!s-M=4.0$}
\psfrag{ksm50}{\Large$K\!s-M=5.0$}
\psfrag{ksm60}{\Large$K\!s-M=6.0$}
\psfrag{ksm67}{\Large$K\!s-M=6.7$}
\psfrag{ksm70}{\Large$K\!s-M=7.0$}
\resizebox{0.95\hsize}{!}{\includegraphics[]{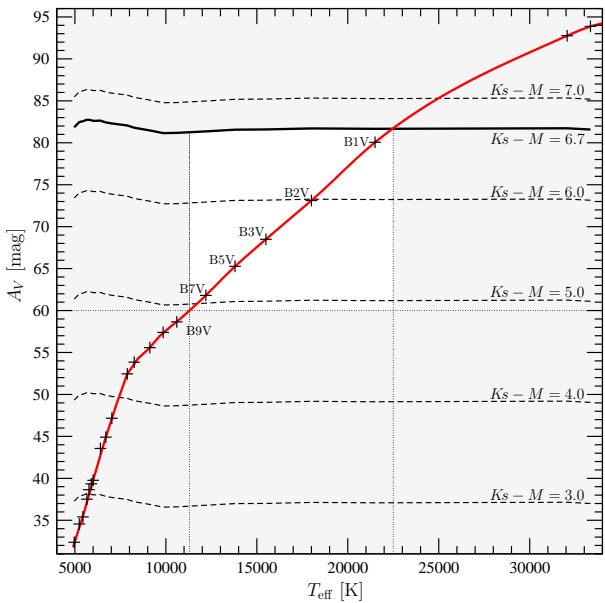}}
\caption{\label{f:ksm}Analysis of infrared colours $K\!s-M$ for combinations of
effective temperatures of stellar photospheres and visual extinctions. The
colours are taken from \citet{ducati01}. The measured upper limit of $K\!s-M
\leq 6.7$ is plotted as a solid line resulting in an upper extinction limit of
$A_V\leq81.7$. The lower extinction limit of $A_V \geq 60$ is denoted by the
white box. The crosses -- interpolated by a solid line -- demonstrate the valid
range for a reddened main-sequence star of $K\!s = 19.28$ at a distance of
2.1\,kpc, neglecting excess emission. The temperatures and spectral types are
taken from \citet{blum00}. The resulting valid range determined by $T_{\rm eff}
= 11300 \ldots 22500$\,K is highlighted by the white box.}
\end{figure}

In order to address this question, we analysed the NIR colours derived from the
values and upper limits given in Tab.~\ref{t:phot}. Assuming the $K\!s-M$ colour
is entirely due to a reddened photosphere, the measured upper limit of
\mbox{$K\!s-M \leq 6.7$} can be used to set an upper limit on the
mass of the protostar. Fig.~\ref{f:ksm} displays the parameter space of $K\!s-M$
colours for given combinations of the visual extinction and the effective
temperature of a main-sequence star. The colours were derived from
\citet{ducati01}, and the temperatures are assigned to spectral types and
stellar masses according to \citet{blum00}. The loci of a reddened main-sequence
star with an apparent magnitude of $K\!s=19.28$ is indicated by crosses,
interpolated by a solid line. This diagram yields the following two results:
First, for the previously determined lower extinction limit of $A_V=60$ the
$K\!s-M$ colour attained by extinction alone amounts to 4.9. Hence, any
additional NIR colour excess must be less than $K\!s-M=1.8$. Second, the
measured $K\!s-M$ colour determines the upper extinction limit to $A_V\leq81.7$.
This constrains also the effective temperature to $T_{\rm eff} \leq 22500$\,K
which is equivalent to a spectral type just above B1, being in the mass range of
about $8\,M_{\sun}$ for a main-sequence star \citep{blum00}. In this way, we
corroborate the upper mass limit given by \citet{sako05}. At this point, we want
to emphasise that not only the colours but also the photometry and upper limits
in Tab.~\ref{t:phot} are consistent with these estimates.

\subsection{The jet emission}
\label{s:jet}

The collimated H$_2$ jet detected by \citet{nuernberger07} originates at the
disc centre and extends as far as 3\farcs6 to the southwest. The SINFONI data
suggest a blue-shifted motion. Correcting for the disc inclination of $12^\circ$
\citep{steinacker06}, a protostellar jet with a typical velocity of
100\,km\,s$^{-1}$ \citep{reipurth01} would reach the current terminus within
365~years. The expected northeastern counter-jet was not detected by
\citet{nuernberger07}.


The NACO $K\!s$ image (Fig.~\ref{f:hk_image}, right panel) gives a first
hint for the existence of such a feature, because the position angle between the
point source and the jet-like feature extending to the northeast agrees with the
direction of the H$_2$ jet. This alignment indicates that we either see the
origin of the counter-jet or at least circumstellar matter entrained and heated
by its interaction. The northeastern extension of the central H$_2$ detection by
SINFONI that coincides with the jet-like feature may be an indication for
residual jet emission. Therefore, we suggest that the counter-jet might exist,
but it is strongly extinguished by the circumstellar material. This agrees with
the picture suggested by \citet{steinacker06} that the near side of the disc is
inclined to the northeast and partly covers the latitudes above the disc plane.
On the other hand, this jet may fail to produce detectable shocks due to a lack
of dense material that has been cleared by the wind and radiation of the cluster
OB stars that are mostly located on the northeastern side of the disc.

\subsection{The bipolar nebula}
\label{s:jet_main}

The accretion disk is associated with a bipolar hourglass shaped structure
\citep{chini04nat}. The northeastern cavity hosts an emission nebula that
resembles the working surface of protostellar jets as seen in other outflows
like HH\,47 \citep{heathcote96,hartigan05}. Our new NACO $H$~band image shows a
similar, but fainter feature inside the southwestern cavity. In the left panel
of Fig.~\ref{f:hk_image}, we have superimposed the contours of the northeastern
emission onto the southwestern nebula and vice versa after rotating them by
$180^\circ$. Besides a similar morphology, there is a symmetry of the peak
emission with respect to the central point-source. Both nebulae have a projected
distance of 1500\,AU from the central object and their origins are located along
an axis that matches the position angle of the star/jet system.

Using its full size at half maximum in the $H$~band, the northeastern nebula
covers an opening angle of $\sim 70^\circ$ with respect to the central
point-source; it has a projected length of 2500\,AU (1\farcs2) and a projected
breadth of 840\,AU (0\farcs4). In the same area, the NACO data attain
$J=17.59$\,mag$/\sq''$, $H=16.68$\,mag$/\sq''$ and $K\!s=16.80$\,mag$/\sq''$
with the brightness peaking along the assumed jet direction. We find an $H$~band
excess of $>\!1.2\,{\rm mag}/\sq''$ to what would be expected for the thermal
emission of a black~body derived from the $JHK$ data. This can be explained
either by a combination of scattering and extinction or by line emission.

Radiation from singly ionised iron -- together with H$_2$ -- is the main coolant
of shocked gas in the NIR \citep{caratti06}. [Fe\,{\sc ii}] lines are commonly
observed in J shocks, where the H$_2$ is completely dissociated
\citep{hollenbach89,kumar03}. Therefore, it is a good candidate for
contaminating the $H$~band continuum flux. However, apart from a spurious
detection of [Fe\,{\sc ii}] emission ahead of the northeastern nebula, in our
NTT/SOFI narrow~band data (not shown) it is strongly concentrated towards the
disc centre. This seems to be a typical phenomenon for many protostellar jets
\citep{itoh01,davis06}. Therefore, the origin of the strong $H$~band emission is
probably not caused by [Fe\,{\sc ii}].

Another more likely explanation is that the nebulae are scattered light reddened
by extinction from the disc material. The spectroscopic accretion signatures
\citep{chini04} imply that most of the scattered photons originate from the
disc centre. \citet{stark06} present Monte Carlo radiative transfer models
of circumstellar discs around young stellar objects. They show that NIR colours
of the disc centres affected by light scattering and extinction strongly depend
on the infall rate, the inclination angle, and the cavity opening angle of the
disc. Their models cover infall rates up to $10^{-5}\,M_{\sun}$\,yr$^{-1}$
producing exceptionally blue colours of $H-K=0.6$ for nearly edge-on discs.
Assuming that the accretion rate of the large M\,17 disc of about
$10^{-4}\,M_{\sun}$\,yr$^{-1}$ \citep{nuernberger07} is of the same order as the
envelope infall rate, we would expect even bluer colours which is at least
qualitatively consistent with our result of $H-K\!s=0.12$.

\subsection{H\boldmath{$_2$} emission at the disc surface}

In the right panel of Fig.~\ref{f:hk_image}, we have superimposed contours of
the SINFONI data of \citet{nuernberger07} to compare the highly resolved H$_2$
$1-0$\,S(1) emission with the $K\!s$ continuum. Apart from the prominent
emission from the disc centre and the southwestern jet, there is also molecular
hydrogen emission at the position of the northwestern disc lane and -- although
much weaker -- at the northeastern nebula. As the H$_2$ emission in the disc
seems to peak at its surface, an external excitation process like UV pumping
might be in action. Indeed, PDRs (photo-dissociation regions) are known to
possess a fair amount of molecular hydrogen just below its surface that can be
excited by FUV (far ultraviolet, $6\,{\rm eV} < h\nu < 13.6$\,eV) radiation
\citep{tielens85}. The existence of such a PDR is also supported by the
detection of emission in the Spitzer 8.0\,$\umu$m band that includes PAH bands
(polycyclic aromatic hydrocarbon molecules).

Two possible sources for the exciting UV photons have to be considered: (a) the
surrounding cluster stars, and (b) the embedded star in the disc centre itself.
If excited by the surrounding cluster stars, one would expect the entire disc
surface to be a PDR that would be detected by H$_2$ line emission. However,
because the size of the PDR is not only a function of the temperature but also
of the density distribution inside the disc, an isotropic UV radiation field
does not necessarily result in a uniform PDR. The major emission occurs at the
northeastern side of the disc facing the hottest cluster stars. If the PDR is
due to the embedded central object, a detailed analysis of the excitation
process in the disc could help to further constrain the nature of the forming
star.

\section{Conclusions}

With the new observations, we have further clarified the nature of the emission
in the centre of the large M\,17 disc and its associated nebula in the following
aspects:

\begin{enumerate}
\item The faint central elliptical emission is resolved for the first time into
a point-source and a tail that extends to the northeast. This morphology
suggests a single protostar accompanied by a jet.

\item The jet-like feature is headed in the opposite direction of a collimated
H$_2$ jet.

\item A southwestern emission nebula is turning the outflow pattern from the
disc centre into a fairly symmetric morphology. The intensity ratio of the
bipolar emission, however, is highly asymmetric, probably due to the inclination
of the dusty disc.

\item The bipolar nebula might be scattered light which is
reddened by extinction caused by the disc material.

\item Various estimates of the extinction towards the embedded protostar yields
a range of $60\leq A_V \leq 82$.

\item The observed 2.2\,$\umu$m flux is equivalent to a protostar with a mass
range of $2.8\,M_{\sun} \leq M_\star \leq 8\,M_{\sun}$.
\end{enumerate}

The new data strengthen the evidence for an intermediate to high-mass
protostellar object being formed by accretion.

\section*{Acknowledgements}
We thank the VLT team for performing the {\jhklm} observations in service mode.
This work was partly funded by the Nordrhein--Westf\"alische Akademie der
Wissenschaften. M.~Nielbock acknowledges the support by the Deutsche
Forschungsgemeinschaft, DFG project SFB~591 and the Ministerium f\"ur
Innovation, Wissenschaft, Forschung und Technik (MIWFT) des Landes
Nordrhein--Westfalen. We thank M.~Haas for helpful discussions. The constructive
respond of the anonymous referee is highly appreciated.

This publication makes in part use of data products from the Two Micron All Sky
Survey, which is a joint project of the University of Massachusetts and the
Infrared Processing and Analysis Center/California Institute of Technology,
funded by the NASA and the National Science Foundation.

This work is based in part on observations made with the Spitzer Space
Telescope, which is operated by the Jet Propulsion Laboratory, California
Institute of Technology under a contract with NASA.

\bibliographystyle{mn2e}
\bibliography{references}

\bsp
\label{lastpage}
\end{document}